\documentclass[sigconf, screen]{acmart}
\AtBeginDocument{%
  }

\setcopyright{acmlicensed}
\copyrightyear{2018}
\acmYear{2018}
\acmDOI{XXXXXXX.XXXXXXX}
\acmConference[Conference acronym 'XX]{Make sure to enter the correct
  conference title from your rights confirmation email}{June 03--05,
  2018}{Woodstock, NY}
\acmISBN{978-1-4503-XXXX-X/2018/06}

\usepackage{array}
\usepackage{longtable}
\usepackage{adjustbox}
\usepackage{graphicx}
\usepackage{booktabs}
\usepackage{array}
\usepackage{geometry}
\usepackage{booktabs,tabularx,array,ragged2e,xcolor,hyperref}
\usepackage[nameinlink]{cleveref} 

\newcolumntype{L}{>{\RaggedRight\arraybackslash}p{0.27\textwidth}}
\newcolumntype{Y}{>{\RaggedRight\arraybackslash}X}
\newcommand{\thickrule}{\specialrule{1.2pt}{0pt}{2pt}}
\newcolumntype{C}[1]{>{\centering\arraybackslash}p{#1}}
\begin{document}

\title["Not Just Me and My To-Do List": Understanding Challenges of Task Management \\ for Adults with ADHD and the Need for AI-Augmented Social Scaffolds]{"Not Just Me and My To-Do List": Understanding Challenges of Task Management for Adults with ADHD and the Need for AI-Augmented Social Scaffolds}

\author{Jingruo Chen}
\authornote{Both authors contributed equally to this research.}
\email{jc3564@cornell.edu}
\affiliation{%
  \institution{Cornell University}
  \city{Ithaca}
  \state{New York}
  \country{United States}}

\author{Yibo Meng}
\authornotemark[1]
\email{mengyb22@mails.tsinghua.edu.cn}
\affiliation{%
  \institution{Tsinghua University}
  \city{Beijing}
  \country{China}
}

\author{Kexin Nie}
\email{niekexinbella@gmail.com}
\affiliation{%
  \institution{The University of Sydney}
  \city{Sydney}
  \country{Australia}
}

\renewcommand{\shortauthors}{Chen et al.}

\begin{abstract}
Adults with ADHD often face challenges with task management, not due to a lack of willpower, but because of emotional and relational misalignments between cognitive needs and normative infrastructures. Existing productivity tools, designed for neurotypical users, often assume consistent self-regulation and linear time, overlooking these differences. We conducted 22 semi-structured interviews with ADHD-identifying adults, exploring their challenges in task management and their coping mechanisms through socially and emotionally scaffolded strategies. Building on these insights, we conducted a follow-up speed dating study with 20 additional ADHD-identifying adults, focusing on 13 speculative design concepts that leverage AI for task support. Our findings reveal that task management among adults with ADHD is relationally and affectively co-constructed, rather than an isolated individual act. Overall, we provide (1) empirical insights into distributed and emotionally scaffolded task management practices, (2) design implications for socially-aware AI systems that support co-regulation and nonlinear attention rhythms, and (3) an analysis of user preferences for different AI design concepts, clarifying which features were most valued and why.
\end{abstract}



\received{20 February 2007}
\received[revised]{12 March 2009}
\received[accepted]{5 June 2009}

\begin{teaserfigure}
    \centering
  \includegraphics[width=\textwidth]{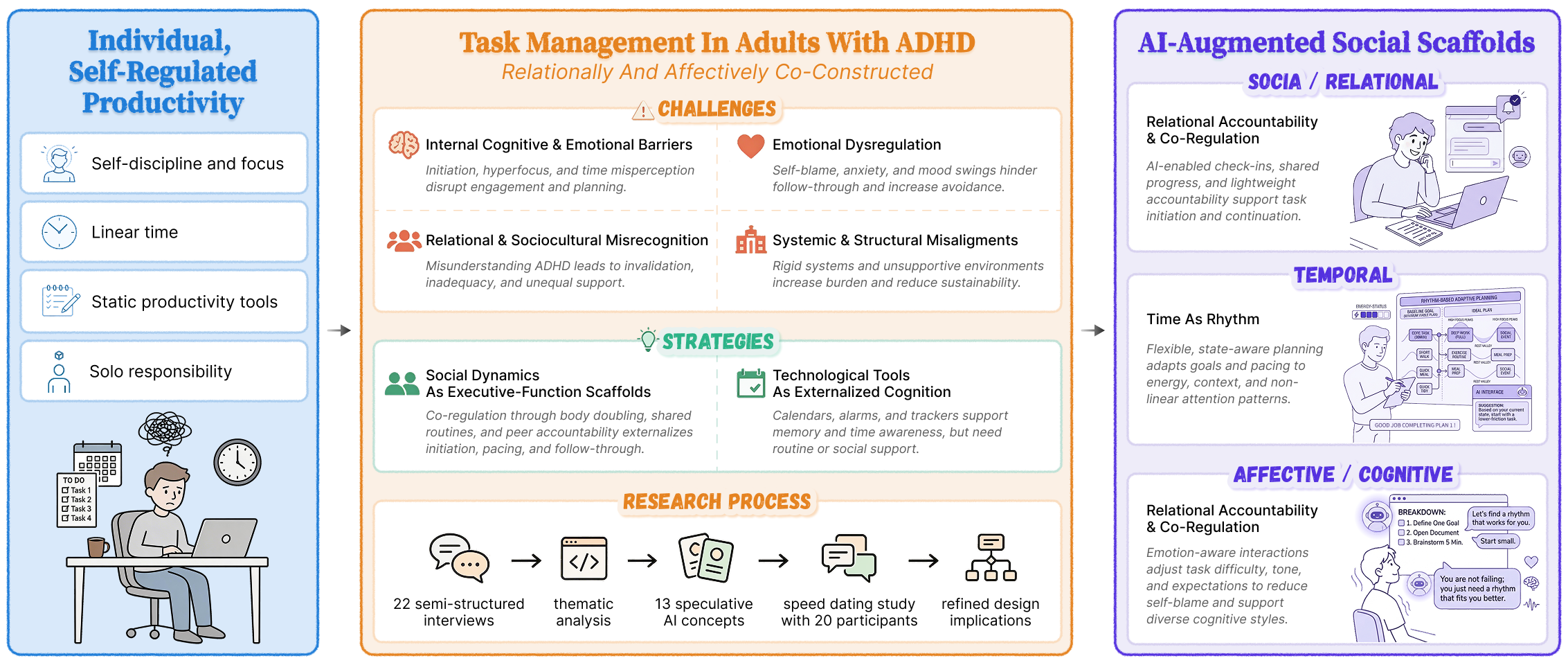}
  \caption{Conceptual overview of task management in adults with ADHD as a relationally and affectively co-constructed process.}
  \Description{Conceptual overview of task management in adults with ADHD as a relationally and affectively co-constructed process.}
  \label{fig:teaser}
\end{teaserfigure}
\maketitle

\section{Introduction}
Attention Deficit Hyperactivity Disorder (ADHD) is a neurodevelopmental condition characterized by persistent patterns of inattention, hyperactivity, and impulsivity that extend beyond childhood into adulthood  \cite{wender2001adults}. In adults, ADHD manifests as a complex interplay of cognitive, emotional, and relational challenges \cite{kritika2025ultimately}, including difficulties in initiating tasks, sustaining focus, managing time, regulating emotions, and maintaining motivation \cite{barkley1998attention}. These challenges are not merely a matter of willpower but are deeply rooted in neurocognitive differences that affect how individuals perceive, prioritize, and execute tasks \cite{boonstra2005executive}.

Research has consistently shown that ADHD is associated with executive dysfunction, a core deficit that impairs an individual's ability to organize, plan, and initiate actions \cite{adamou2013occupational}. For adults with ADHD, task management is rarely an isolated or purely cognitive process. Instead, it is relationally and affectively co-constructed, relying heavily on external supports, social scaffolding, and adaptive routines\cite{eagle2023proposing, Burtscher2024work}. Social interactions, emotional regulation, and supportive technologies play critical roles in compensating for these cognitive challenges, transforming task management into a socially and emotionally scaffolded practice \cite{Eagle2024Body}.

Despite this, most productivity tools and AI-based task management systems are designed around neurotypical assumptions, including stable attention, linear time perception, and individual self-regulation \cite{campbell2023adhd, 10.1145/3750069.3750114}. Empirical studies show that commonly used tools, such as calendars, to-do lists, planners, and reminder systems, are perceived as significantly less effective by adults with ADHD than by non-ADHD users, despite similar patterns of adoption and use \cite{desrochers2019evaluation}. Analyses of popular productivity platforms further suggest that these systems prioritize task completion and efficiency while offering limited support for task initiation, time estimation, or emotional regulation \cite{oliveira2025personalized}. This mismatch can contribute to frustration, disengagement, and feelings of inadequacy among adults with ADHD \cite{campbell2023adhd, mette2023time}.

Recent HCI and CSCW research has begun to challenge these limitations by advocating for neurodiversity-oriented and participatory design approaches \cite{Tobias2016Framework, Le2024Iam, spiel2022adhd}. Studies have explored affect-aware systems, social and relational scaffolds (e.g., body doubling, socially assistive robots), and user-driven appropriations of general-purpose technologies, highlighting the importance of co-regulation and customization \cite{Eagle2024Body, O'Connell2024robot, 10.1145/3597638.3608400, 10.1145/3764687.3764713}. However, much of this work remains descriptive, child-focused, or scoped to specific interaction contexts, with relatively few studies examining how adults with ADHD manage everyday tasks or how designs of AI systems might support these practices in real-world settings \cite{Burtscher2024work, zhu2025focusview}.

To better understand and support the task-management practices of adults with ADHD, we conducted semi-structured interviews with 22 ADHD-identifying adults, examining their everyday challenges, coping strategies, and perceptions of support. Building on these insights, we then ran a follow-up speed-dating study with 20 additional participants, presenting 13 speculative AI-supported design concepts for task support. This iterative mixed approach surfaced user preferences, clarified trade-offs (e.g., flexibility vs.\ accountability), and refined the concept space.

Our study is guided by three core research questions:
\begin{itemize}
\item \textbf{RQ1:} What challenges do adults with ADHD report in everyday task management, what strategies do they use, and what expectations do they have for supportive systems?
\item \textbf{RQ2:} Given these findings, what design implications and speculative concept space follow for socially and affectively aware task-management support?
\item \textbf{RQ3:} From participants’ reactions to our speculative concepts in speed-dating, what design directions for AI support emerge (including preferences, tensions, and boundaries)?
\end{itemize}

Our findings show that task management for adults with ADHD is relational and affective rather than a purely individual endeavor. Specifically, our contributions are threefold:
\begin{itemize}
\item An empirical account of challenges, strategies, and expectations that structure ADHD task management in everyday life.
\item A synthesis of design implications and a catalog of 13 speculative concepts, indicating which features participants valued and why.
\item Design directions for socially aware, mood-attuned AI systems that support co-regulation, adaptive routines, and non-linear attention rhythms, grounded in speed-dating feedback.
\end{itemize}

Together, these insights argue for rethinking task-management tools for adults with ADHD: moving beyond rigid, individualistic systems toward adaptive, relational, and emotionally intelligent support.

\section{Related Work}
\subsection{Adults with ADHD and Task Management}

Adults with Attention Deficit Hyperactivity Disorder (ADHD) face persistent challenges in task management due to deficits in executive functioning\cite{barkley2002major}, including difficulties with task initiation, sustained attention, organization, and time management \cite{kooij2012adult}. Rather than a simple inability to focus, ADHD is characterized by patterns of inattention alongside episodes of intense overconcentration (“hyper-focus”); the core issue lies in regulating attention when needed rather than an absolute lack of focus \cite{kooij2012adult}. ADHD is increasingly understood as a lifelong condition, with cumulative challenges that persist into adulthood and can substantially impact quality of life and functional capacity \cite{adamis2024functional}. In workplace contexts, ADHD has been linked to higher accident rates and reduced productivity \cite{reynolds1997attention, kessler2009prevalence}.

A common manifestation of these challenges is task procrastination, often driven by inattention, distractibility, and difficulty engaging with complex or monotonous activities \cite{oguchi2021moderating, niermann2014relation}. This pattern is rooted in neurocognitive characteristics of ADHD, including impairments in prospective memory \cite{altgassen2019prospective}. Steel’s Temporal Motivation Theory attributes procrastination to low expectancy, negative task value, and impulsiveness—defined as sensitivity to delayed rewards \cite{steel2007nature}. Empirical work shows that ADHD symptoms are positively associated with task aversiveness and impulsiveness and negatively associated with expectancy, with expectancy and impulsiveness partially mediating the relationship between ADHD and procrastination \cite{netzer2023using}. Emotional dysregulation further compounds these difficulties: adults with ADHD frequently experience heightened frustration or anxiety in response to task demands, reinforcing cycles of avoidance and reduced self-efficacy \cite{adamis2024functional, barkley1998attention, soler2023evidence}.

Contemporary diagnostic frameworks describe three ADHD presentations: predominantly inattentive (ADHD-I), predominantly hyperactive–impulsive (ADHD-HI), and combined (ADHD-C) \cite{soendergaard2016associations, murphy2002young}. ADHD-I is primarily associated with difficulties in sustained attention and task completion, ADHD-HI with restlessness and impulsivity, and ADHD-C with both symptom clusters \cite{sobanski2008subtype, faraone2000attention, schweitzer2006working}. These presentations reflect dominant symptom patterns rather than distinct disorders, and individuals may shift across them over the life course \cite{willcutt2012prevalence, edition1980diagnostic}. Longitudinal studies suggest that hyperactivity and impulsivity often diminish in adulthood, while inattentive symptoms remain more persistent \cite{gibbins2010adhd}. Given fluctuating symptom expression and our mix of clinically diagnosed and self-identified participants (many of whom were unaware of their specific presentation), we focus on shared task-management experiences rather than presentation-stratified analyses.

\subsection{Rethinking Neurodivergent Design in HCI and CSCW}

CSCW and Human-Computer Interaction (HCI) have increasingly recognized the importance of designing systems that accommodate neurodiversity. Traditional design paradigms often assume a normative user model, marginalizing individuals with cognitive differences such as ADHD. In response, recent work has advanced a neurodiversity-oriented perspective that centers the experiences and strengths of neurodivergent individuals \cite{Tobias2016Framework, Le2024Iam}, alongside participatory approaches that position people with ADHD as active stakeholders in research and design \cite{spiel2022adhd}. Philip et al. \cite{Philip2025Social} further argue for a critical disability theory perspective, framing neurodivergent sociality as meaningful rather than deficient.

Reflecting this shift, prior studies demonstrate the value of participatory and co-designed systems for neurodivergent users \cite{O'Connell2024robot}. For example, Stefanidi et al. developed MoodGems, a modular display system supporting emotional expression among children with ADHD through iterative co-design with families and practitioners \cite{Stefanidi2024mood}. Similarly, Pulatova and Kim designed customizable tabletop robots to support fidgeting and sensory regulation for adults with ADHD \cite{Pulatova2024Robot}. Together, these studies emphasize customization, co-design, and affective support, while largely focusing on emotional regulation or sensory engagement rather than everyday task management.

Alongside participatory design, substantial HCI research has targeted attention through individual-level cognitive or perceptual interventions, often focusing on children with ADHD. This includes work on gamification and visual parameters in reading and learning contexts \cite{10.1145/3772068, 10.1145/3675888.3676047, asiry2018extending, phalke2023identification}, as well as AR/VR-based serious games and immersive learning or assessment tools \cite{avila2018towards, 10.1145/3613904.3643021, 10544310, 10.1145/3284497.3284499}. While these systems demonstrate measurable benefits, they are typically short-term or narrowly scoped to specific tasks.

More recently, HCI work has begun to explore social and relational scaffolds for ADHD \cite{ara2025you}. Studies show that college students with ADHD respond positively to socially assistive robots providing non-verbal attention feedback \cite{O'Connell2024robot} or companionship \cite{10.5555/3721488.3721695}, as well as conversational agents that support shared accountability in daily routines \cite{10.1145/3334480.3382948}. In parallel, online platforms and social media function as spaces for peer support, validation, and identity work among adults with ADHD \cite{10.1145/3597638.3608400, leveille2024tell, 10.1145/3778352}. However, these social resources remain largely disconnected from task-management systems and workplace tools.

Taken together, prior work offers rich insight into neurodivergent experiences and coping practices \cite{campbell2023adhd, 10.1145/3663547.3746325}, yet relatively few studies translate these insights into evaluated systems that support everyday task management for adults with ADHD. Calls for greater attention to neurodivergent self-determination in work contexts \cite{Burtscher2024work} and analyses of practices such as body doubling \cite{Eagle2024Body} highlight this gap. Emerging systems like FocusView illustrate the promise of attention customization in specific contexts but remain limited in scope \cite{zhu2025focusview}. Overall, while CSCW and HCI increasingly emphasize participatory, affect-aware, and socially grounded design, support for everyday task management and workplace practices among adults with ADHD remains underexplored.

\subsection{AI in Task Management}

Artificial Intelligence (AI) has become integral to task management systems, offering features such as automated scheduling, reminders, and productivity analytics \cite{10.1145/3663384.3663404, 10.1145/3173574.3173632, 10.1145/3596671.3598572}. However, these systems are often designed around neurotypical behaviors, failing to account for variability in attention, emotional regulation, and cognitive rhythms characteristic of adults with ADHD \cite{10.1145/3750069.3750114, campbell2023adhd}. Analyses of popular productivity tools (e.g., Trello, Todoist, Focus@Will, Forest) show that while they support general organization, they often lack features critical for ADHD users, such as fine-grained task decomposition and personalized time-management strategies, prioritizing task completion over individualized progress \cite{oliveira2025personalized}. Survey research further demonstrates this gap as Desrochers et al. found that adults with ADHD reported significantly lower perceived effectiveness of commonly used tools, despite relying on similar strategies as non-ADHD participants, attributing this disparity to design misalignment rather than lack of use \cite{desrochers2019evaluation}.

Prior work has explored AI-supported task management in broader contexts. Morrison et al. examined AI-powered reminders in collaborative settings, identifying concerns around privacy and notification overload \cite{10.1145/3653701}. Ma and Ren introduced ProactiveAgent, a context-aware system leveraging large language models (LLMs) for personalized reminders \cite{10.1145/3586182.3625115}. While these systems enhance efficiency, they typically assume stable user routines, which may not align with ADHD task-management patterns.

Other research has examined adaptive and supportive designs relevant to neurodivergent needs. Pradhan et al. showed that multimodal AI reminders combining digital and physical cues can enhance accessibility for users with dementia, suggesting applicability to ADHD users facing fluctuating attention and motivation \cite{10.1145/3597638.3614506}. Faulring et al. developed RADAR, an agent-assisted system that learns user strategies to manage email overload, improving task performance \cite{10.1145/1719970.1719980}. Pinto et al. examine how adults with ADHD appropriate general-purpose generative AI tools such as ChatGPT to support executive functioning, emotional regulation, and communication, highlighting the role of user-driven AI adoption in inclusive design \cite{10.1145/3764687.3764713}.

However, AI-driven task management also raises challenges. Qiao et al. and He et al. identify risks of overreliance on automation and loss of user agency \cite{10.1145/3706598.3714103, 10.1145/3596671.3598572}, which may be amplified for ADHD users who oscillate between hyperfocus and inaction. Time perception further complicates task management: ADHD users often struggle with estimating task duration and maintaining temporal awareness \cite{mette2023time, ptacek2019clinical}. Ahmetoglu et al. link these difficulties to the planning fallacy, noting that although mitigation strategies exist, commercial systems rarely implement them effectively \cite{10.1145/3663384.3663404}.

In summary, while AI shows promise for task management, existing systems often lack the flexibility needed to support adults with ADHD. Our study builds on this work by examining how socially and emotionally aware AI can better support neurodivergent task management.

\section{Method}

\begin{figure*}[t]
  \centering
  \includegraphics[width=\textwidth]{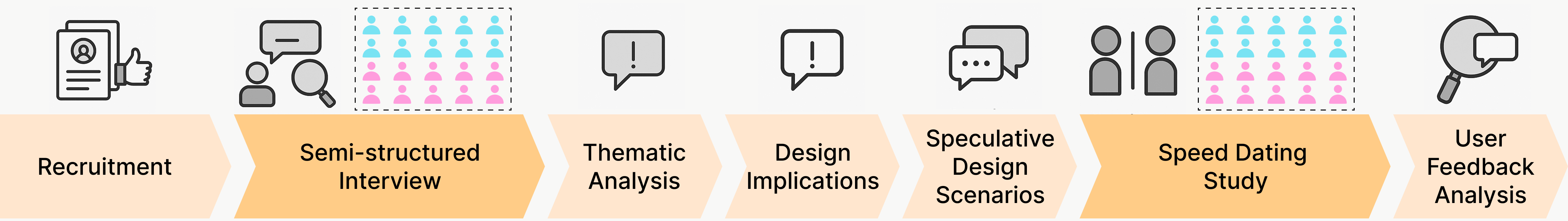}
  \caption{Method Overview}
  \label{figure: Method}
\end{figure*}

To explore how adults with ADHD manage tasks and imagine potential AI support, we conducted 22 \textbf{semi-structured interviews} with individuals who either had a formal diagnosis of ADHD or strongly self-identified as having ADHD (P1–P22). Participants were recruited through social media, peer referrals, and university outreach. Interviews lasted between 25 and 35 minutes; for interviews exceeding 25 minutes, we included a short five-minute break to support participant focus and comfort. All interviews were conducted remotely and audio-recorded with participants’ consent. Each participant received \$40 as compensation. Our study protocol was approved by our institution’s ethics review board.

Building on insights from these interviews, we developed speculative design scenarios illustrating potential ways AI systems could support ADHD-related challenges, such as task initiation, time management, and emotional regulation (Figure \ref{figure: Method}). We then conducted a follow-up \textbf{speed dating study} with 20 additional adult ADHD participants (S1–S20), none of whom had participated in the interview phase. Each session lasted approximately 25 minutes and consisted of a series of short, structured engagements with our speculative AI concepts. Participants were encouraged to reflect candidly on their reactions, both practical and emotional. All interviews and speed dating sessions were held remotely, lasting 25 minutes, and were also recorded with consent.

\subsection{Participants}
Our interview participants (ID P1–P22) came from diverse educational and professional backgrounds and self-identified as adults with ADHD (Table \ref{tab:interview_demographic}). Participants were asked to reflect on their personal experiences with planning, task execution, attention, emotional regulation, and their use (or rejection) of various digital tools. While a fourth of the participants had prior experience with ADHD medication, most managed their symptoms without pharmacological support, often drawing on self-developed coping strategies or informal social networks.

The 20 participants in our follow-up speed dating study (S1–S20) did not overlap with the initial group. They were recruited with the intention of ensuring a range of experiences with AI tools. All speed dating sessions used the same set of speculative design artifacts and were analyzed for thematic feedback.

\subsection{Thematic Analysis}
We conducted a thematic analysis of interview transcripts using a hybrid inductive coding approach \cite{clarke2017thematic}. Coding was performed by two researchers, who regularly met to discuss emerging codes and resolve discrepancies. Themes were iteratively refined to ensure they reflected the lived experiences of participants rather than preconceived assumptions. The same process was used to analyze speed dating feedback, with particular attention to participants’ emotional reactions to the AI scenarios.

\subsection{Positionality}
Our research team includes members with and without lived experience of ADHD. This positional diversity informed both the design of the interview and speed dating protocols and the interpretation of participant narratives. While we recognize that our work reflects our shared academic background and interest in human-AI interaction, we intentionally centered the voices of participants in shaping both the speculative concepts and the resulting analysis, and included participants with diverse experiences in the usage of AI.

\section{Interview Findings}

\begin{table*}[t]
\centering
\small
\setlength{\tabcolsep}{5pt}
\renewcommand{\arraystretch}{1.1}

\begin{tabular}{l c c l l c p{7.5cm}}
\toprule
\textbf{ID} & \textbf{Gender} & \textbf{Age} & \textbf{Occupation} & \textbf{Education} & \textbf{Diag.\textsuperscript{†}} & \textbf{Tools Used} \\
\hline
P1  & M & 19 & Student        & Bachelor’s    & CD  & Todoist; Trello; Forest app; digital timer \\
P2  & F & 19 & Student        & Bachelor’s    & SSI & Reminders; sticky notes \\
P3  & F & 18 & Student        & Bachelor’s    & CD  & None \\
P4  & M & 38 & Teacher        & PhD           & CD  & None \\
P5  & F & 22 & Student        & Bachelor’s    & CD  & Forest app; digital timer \\
P6  & M & 32 & Office worker  & Bachelor’s    & CD  & Forest app; to-do list \\
P7  & F & 18 & Student        & Bachelor’s    & CD  & Sticky notes; notebooks \\
P8  & F & 34 & Doctor         & PhD           & CD  & Planner app; Forest app; digital timer \\
P9  & M & 23 & Teacher        & PhD           & SSI & Digital reminders; to-do list \\
P10 & F & 33 & Manager        & High school   & CD  & To-do list \\
P11 & F & 33 & Teacher        & Bachelor’s    & CD  & To-do list \\
P12 & F & 18 & Student        & Bachelor’s    & SSI & None \\
P13 & M & 45 & Office worker  & Bachelor’s    & CD  & None \\
P14 & F & 19 & Student        & Bachelor’s    & CD  & Reminders \\
P15 & M & 18 & Student        & High school   & CD  & Planner apps \\
P16 & F & 44 & Office worker  & Middle school & SSI & None \\
P17 & M & 22 & Office worker  & Middle school & CD  & Planner apps \\
P18 & F & 55 & Farmer         & Elementary    & CD  & Sticky notes \\
P19 & M & 23 & Office worker  & Middle school & SSI & None \\
P20 & F & 55 & Unemployed     & None          & CD  & Notebooks \\
P21 & M & 22 & Student        & Bachelor’s    & SSI & To-do list \\
P22 & F & 22 & Student        & Master’s      & CD  & Reminders \\
\hline\hline
\end{tabular}

\caption{Demographic information of semi-structured interview participants (self-reported).}
\label{tab:interview_demographic}

\par\smallskip
{\small \emph{Note.} \textsuperscript{†}\,Diag.\ uses abbreviations for Diagnosis: CD = clinically diagnosed; SSI = strongly self-identified. Total clinically diagnosed: 16; Total strongly self-identified: 6.}
\end{table*}

\begin{figure*}[t]
  \centering
  \includegraphics[width=\textwidth]{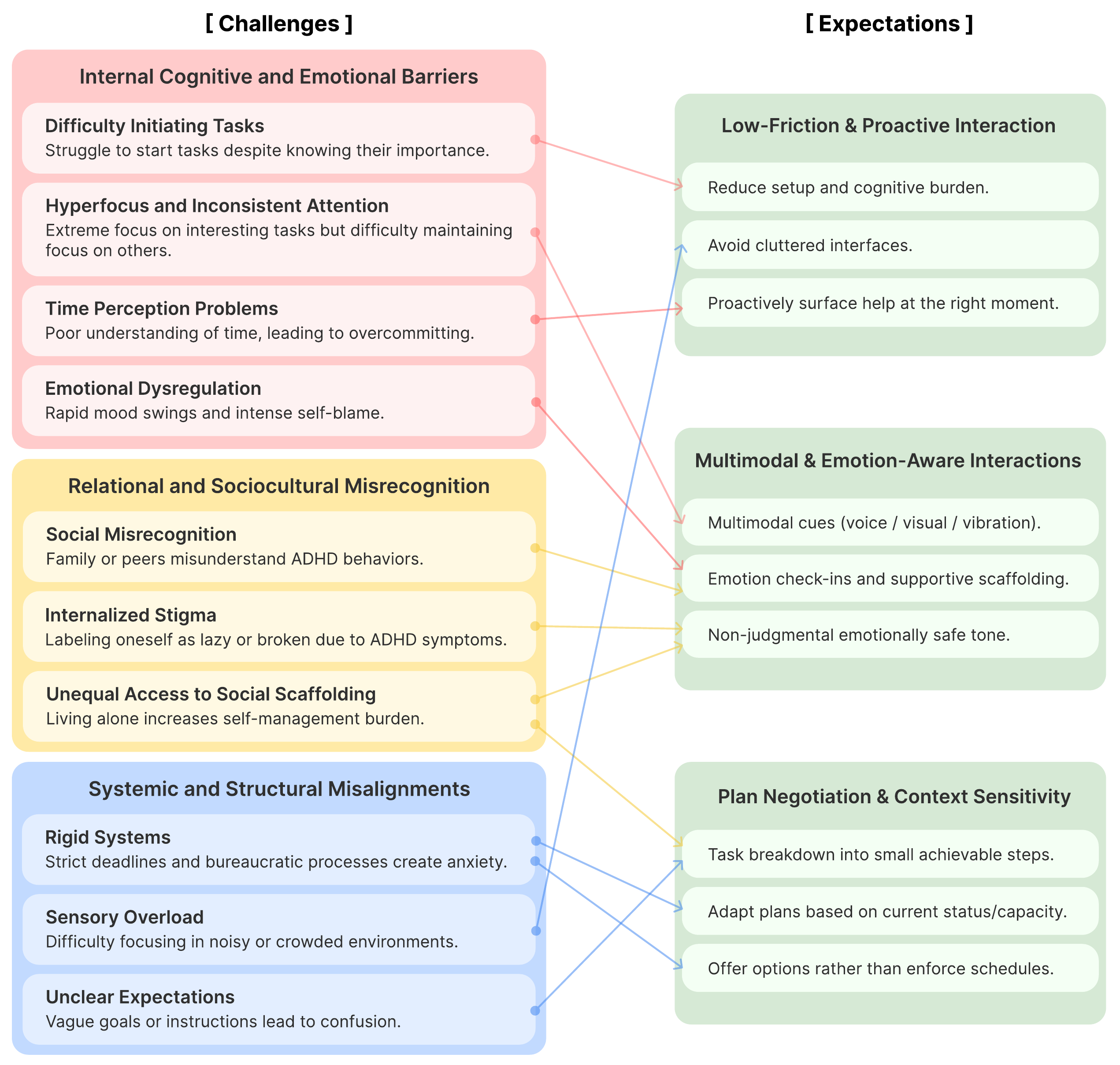}
  \caption{Challenge–Expectation mapping from interview findings}
  \label{interview_findings}
\end{figure*}

Our analysis of 22 interviews reveals that task management among adults with ADHD is shaped by a constellation of internal struggles, systemic mismatches, and compensatory strategies. To capture this complexity while maintaining narrative clarity, we organize our findings into three overarching sections (see Table \ref{tab:adhd-summary-findings}): (1) challenges (section \ref{section: internal challenges}, section \ref{section: relational challenges}, and section \ref{section: systemic challenges}), (2) strategies (section \ref{section: social strategies} and section \ref{section: technological tool strategies}), and (3) expectations (section \ref{section: AI expectation}). This structure allows us to first understand the tensions participants face, then explore how they cope, and conclude with what kinds of support they envision from AI systems (see Figure \ref{interview_findings}).

\subsection{Challenges: Internal Cognitive and Emotional Barriers}
\label{section: internal challenges}
\subsubsection{Executive Dysfunction and Temporal Disorientation}
\label{section: Executive Dysfunction and Temporal Disorientation}
\textbf{Task Initiation Paralysis and Psychological Inertia.}
Almost all participants reported profound difficulty initiating tasks, even when they understood them or recognized their importance (P1, P5, P6, P17, P19). This "task paralysis" extended beyond professional and academic responsibilities to routine activities such as getting out of bed, brushing teeth, or sending a simple message. For several individuals, this inability to initiate was described as a kind of psychological inertia: being mentally aware of the need to act but physically unable to start. Participants emphasized that their struggle was not with comprehension or motivation in abstract terms, but with the practical transition into action.

As some participants described their entire productivity hinging on whether they could bypass this initial block (P6, P19), the phenomenon was reported as particularly distressing because it undermined participants’ confidence in their own autonomy and created cycles of shame and avoidance. One participant described: “\textit{I obviously want to brush my teeth and wash my face, but I stay in bed all the time; there will be a big start-up difficulty.}” (P17).

\textbf{Hyperfocus and Flow States.}
\label{Hyperfocus and Flow States}
Participants’ ability to sustain attention and initiate tasks was highly dependent on perceived relevance and intrinsic interest. Engagement was often described as binary, either hyperfocus or total disengagement, with few productive middle states. Several participants reported experiencing “flow-like” states when working on tasks they found meaningful or enjoyable (P1, P6, P13). In these moments, attention and motivation aligned, and participants were able to work for extended periods without distraction. P13 recalled, \textit{“When it’s something I really like, I can write for ten hours straight, and it feels effortless.”} However, these states were unpredictable and often not replicable on demand.

\textbf{Time Perception Disorders and Unrealistic Planning Cycles.}
A large subset of participants described fundamental misalignments in their perception of time, particularly when attempting to plan for tasks (P1, P5, P10, P14). Many participants reported consistently overestimating what they could accomplish in a day, in part because they based plans on rare episodes of hyperfocus, which they mistakenly assumed were replicable (P1, P14). As a result, planning became not only aspirational but also emotionally taxing as plans were often written with enthusiasm, but then abandoned, leading to cycles of disappointment and self-blame (P10). Inaccurate time estimation also contributed to an aversion to planning. Some participants noted that unless a deadline was immediate and tangible, their perception of its urgency remained low (P5). Time became a vague, intangible element of their environment unless artificially made visible by alarms, friends, or crises. As one participant put it: “\textit{Unless I pay special attention, my perception of time is as if I want to deliberately forget it.}” (P5).

\textbf{Deadline Addiction and Crisis-Driven Productivity.}
Despite the emotional toll, many participants reported being most productive under crisis conditions, particularly when faced with a looming deadline (P1, P3, P15). This behavior was not simply a matter of last-minute effort but had become an ingrained coping strategy, often described with a mix of frustration and dependency. Several participants admitted to "needing" the emotional spike of deadline pressure to mobilize effort (P1, P15), with some associating this pattern with a form of adrenaline addiction.

While some participants had developed intricate rituals around these last-minute bursts, including pre-deadline planning notebooks and crisis-mode schedules, others reported using medication as the only way to sustain the prolonged productivity required to meet deadlines (P3). Across the board, however, most acknowledged that this crisis-driven approach was unsustainable, physically exhausting, and damaging to long-term mental health. As P1 explained, \textit{“I feel that I have formed a kind of satisfaction [from] this kind of dependence on pressure.”}

\subsubsection{Emotional Dysregulation}
\label{section: Emotional Dysregulation}
In addition to cognitive and temporal challenges, participants described complex emotional experiences that shaped their approach to task management, social relationships, and self-perception.

\textbf{Emotional Avoidance of Low-Meaning Tasks.}
Tasks that lacked personal relevance were not simply boring but emotionally aversive (P4, P8, P12). Participants described a visceral rejection of these tasks, often accompanied by guilt and frustration. Several noted that traditional motivational strategies (e.g., rewards, timers) were ineffective unless the task had some perceived value. This sensitivity to meaning complicates task management in academic and professional settings where not all obligations are felt intrinsically worthwhile.

\textbf{Emotional Fluctuations and Self-blame.}
Many participants reported experiencing rapid and destabilizing mood shifts throughout the day, often triggered by minor setbacks or interpersonal tensions (P1, P4, P9, P17). These emotional fluctuations frequently led to an inability to complete tasks, regardless of their initial motivation or planning. Some described how a single comment or perceived failure in the morning could derail their entire day’s productivity (P4, P9).

This emotional reactivity is often intertwined with self-blame. Participants reported a tendency to harshly criticize themselves for even momentary lapses in concentration, which in turn compounded stress and reduced focus further (P17). One participant reflected: \textit{“I often blame myself for a moment of distraction and the originally thought-out plan keeps getting tangled in my brain.”} (P17).

\textbf{Fear of Sustainability and Self-Sabotage.}
Several participants described a paradoxical relationship with success. While they aspired to achieve, they also feared the responsibility of maintaining success, which often triggered self-sabotaging behaviors (P1, P6). For these individuals, moments of progress could provoke anxiety about whether they could sustain their performance, leading them to abandon efforts prematurely. P1 described a pattern of intentionally destroying opportunities to avoid the pressure of maintaining them: \textit{“If I really succeed, it will be very hard to maintain. I might as well smash this possibility with my own hands.”} 

\textbf{Planning Anxiety.} Rigid timelines, bureaucratic paperwork, and fixed expectations often triggered a cycle of anticipatory failure and planning aversion. P14 described a common emotional spiral: \textit{"The plan is always written very well, but I know I can't do it."} Likewise, P10 shared: \textit{"Every time I write a plan, I get anxious because I know I won't follow it."} Rather than providing clarity, formalized systems of planning became sources of shame and paralysis.

\begin{table*}[t]
\centering
\caption{Summary of challenges and strategies}
\label{tab:adhd-summary-findings}
\small
\renewcommand{\arraystretch}{1.18}
\begin{tabularx}{\textwidth}{L Y}
\thickrule
\multicolumn{2}{l}{\textbf{RQ1: What are the core challenges of task management for adults with ADHD, and what strategies do they use to cope?}}\\
\thickrule
\textbf{Section} & \textbf{Summary}\\
\midrule

\multicolumn{2}{l}{\textbf{\ref{section: internal challenges} Challenges: Internal cognitive \& emotional barriers}}\\[2pt]

\textbf{Task initiation paralysis \& psychological inertia} &
Persistent difficulty starting even simple actions despite intention; sense of “knowing but not starting” that undermines autonomy and cascades into avoidance.\\

\textbf{Hyperfocus \& flow state} &
Binary engagement pattern (deep absorption vs.\ total disengagement) driven by perceived meaning; productive states are powerful but unpredictable and hard to summon on demand.\\

\textbf{Time perception disorders \& unrealistic planning cycles} &
Chronic misestimation and low felt urgency unless time is made concrete; plans overfit to rare hyperfocus days, producing repeated overcommitment and collapse.\\

\textbf{Deadline addiction \& crisis-driven productivity} &
Reliance on last-minute pressure to mobilize effort; short-term spikes of productivity paired with exhaustion and longer-term instability.\\

\textbf{Emotional avoidance of low-meaning tasks} &
Tasks lacking personal relevance are experienced as emotionally aversive, often triggering visceral resistance, guilt, and frustration. \\

\textbf{Emotional fluctuations \& self-blame} &
Mood shifts and sensitivity to setbacks disrupt progress; reactivity and self-criticism drive cycles of stress and disengagement.\\

\textbf{Fear of sustainability \& self-sabotage} &
Progress can trigger anxiety about maintaining performance, leading to avoidance or intentional disengagement; success itself becomes a source of pressure that undermines continued effort.\\

\textbf{Planning anxiety} &
Rigid planning structures and formalized systems provoke anticipatory failure and anxiety.\\
\thickrule
\multicolumn{2}{l}{\textbf{\ref{section: relational challenges} Challenges: Relational \& sociocultural misrecognition}}\\[2pt]

\textbf{Social misrecognition \& family invalidation} &
ADHD needs are minimized or misunderstood (e.g., “just try harder”), straining relationships and impeding accommodation.\\

\textbf{Internalized inadequacy} &
Deficit narratives are absorbed as laziness or moral failure; long-standing labels from school/work shape self-concept even post-diagnosis.\\

\textbf{Resistance \& reframing} & Effort to construct an empowered ADHD identity can be emotionally taxing.\\

\textbf{Unequal access to social scaffolding} &  Living alone or in unsupportive environments increased the burden of self-management. \\

\thickrule
\multicolumn{2}{l}{\textbf{\ref{section: systemic challenges} Challenges: Systemic \& structural misalignments}}\\[2pt]

\textbf{Institutional and environmental misalignment} & Administrative rigidity, notification clutter, and noisy settings amplify cognitive load and undermine the sustainability of work and study.\\

\textbf{Over-reliance \& avoidance} & Tool affordances can enable planning-as-avoidance in the absence of external anchors.\\

\thickrule
\multicolumn{2}{l}{\textbf{\ref{section: social strategies} Strategies: Social dynamics as executive-function prosthetics}}\\[2pt]

\textbf{Co-regulation and borrowed structure} &
Body-doubling, shared routines, and friend-based monitoring to externalize initiation, pacing, and follow-through.\\

\textbf{Peer communities and recognition} &
Participation in ADHD communities (online/offline) for resonance, legitimacy, and emotionally safe disclosure.\\

\thickrule
\multicolumn{2}{l}{\textbf{\ref{section: technological tool strategies} Strategies: Technological tools as externalized cognition}}\\[2pt]

\textbf{Digital Tools and scaffolded memory} &
Calendars, alarms, lists, and habit trackers are used to offload prospective memory and make time visible; often shared to keep salient.\\

\textbf{Emotional symbolism of tools} &
Bullet journaling and list-making support affect regulation and sense-making around attention and identity.\\

\textbf{Simulated companionship \& AI co-regulation} &
Use of chatbots/apps for prompt-and-report rituals and friend-like nudges that create a felt sense of being “seen.”\\

\thickrule
\end{tabularx}
\end{table*}

\subsection{Challenges: Relational and Sociocultural Misrecognition}
\label{section: relational challenges}
\textbf{Social Misrecognition and Family Invalidation.} Many participants recounted long-standing experiences of being misunderstood, particularly by family members and the broader public. Misconceptions about ADHD, especially the assumption that it only entails visible hyperactivity, led to the dismissal of their struggles (P10, P17). As P10 put it, \textit{"Everyone thinks ADHD is only about 'hyperactivity' and can't sit still."} While some participants found emotional support from family, many reported familial responses that minimized, invalidated, or pathologized their behaviors. P6 recalled: \textit{"When I told my sister I had ADHD, she said, 'Then just correct it.'"} Crucially, this invalidation often came not from malice but from ignorance. Participants emphasized that otherwise loving relationships could turn strained when ADHD was brought up, with misunderstandings rooted in generational or cultural blind spots. P19 described their mother conflating ADHD with mere restlessness and projecting her own childhood discipline strategies, failing to grasp the neurological basis of the condition.

\textbf{Internalized Inadequacy.} This persistent misrecognition contributed to internalized stigma. Many participants described interpreting their executive functioning challenges as laziness or moral failure well into adulthood (P1, P10, P14). Even post-diagnosis, several struggled to legitimize their needs fully. P10 admitted: \textit{"I know it’s ADHD, but I still feel like I should just try harder. Other people manage; why can’t I?"} The roots of these narratives often trace back to early school experiences. Labels like “lazy,” “messy,” or “disorganized” haunted participants for years, coloring their self-concept despite academic achievements. As P1 recalled: \textit{"Since elementary school, I’ve been called lazy and impulsive."}

\textbf{Resistance and Reframing.} Other participants worked to reframe ADHD outside of deficit models. For them, ADHD traits, such as divergent thinking or emotional sensitivity, became potential strengths when situated in supportive contexts. P7 stated: \textit{"It’s not that I can’t focus; it’s that I focus differently. The problem is the world doesn’t allow for that."} These reappraisals were often nurtured by peer networks, therapeutic settings, or affirming mentors. Still, the effort to construct an empowered ADHD identity was described as emotionally taxing. For some, receiving a formal ADHD diagnosis offered critical relief and emotional clarity. Diagnosis enabled participants to reinterpret their life stories, not as evidence of personal failure but as the result of misunderstood neurodivergence (P11, P13, P17). P13 described the moment as transformative: \textit{"I was very happy on the day of diagnosis; I finally didn’t have to fight myself anymore."}

\textbf{Unequal Access to Social Scaffolding.} 
Beyond how ADHD was understood or reframed, participants also described uneven access to social support as a persistent relational challenge. Not all participants had access to robust support networks. Those living alone or in unsupportive environments described the emotional burden of self-management in isolation (P8, P14). Asking for help repeatedly felt shameful, and the unpredictability of social reciprocity left some reluctant to rely on others at all. Structural conditions, such as living arrangements, relationship stability, or access to peer communities, often shaped participants' ability as much as personal traits did.

\subsection{Challenges: Systemic and Structural Misalignments}
\label{section: systemic challenges}

\textbf{Institutional and environmental misalignment.} Across school and work, participants encountered environments designed for neurotypical rhythms, structures that often demanded forms of attention, memory, and self-regulation that were especially taxing for those with ADHD. This mismatch created not only inefficiency but also emotional strain. P12 recounted losing their campus card over a dozen times, emphasizing how even basic infrastructure, like ID-dependent access, can impose outsized burdens. P4 nearly missed graduation due to a lack of administrative follow-up, lamenting: \textit{"Why didn’t you just remind me once that I didn’t have enough credits?"} Participants described these systems as rigid and unforgiving, where minor oversights snowballed into major setbacks. The issue was not resistance to structure itself, as many participants valued routine, but rather the absence of adaptive mechanisms that accounted for fluctuating cognitive and emotional states.  Workplaces, too, often lacked sensory or procedural accommodations. P4 left their job due to open-office sensory overload, noting that such environments heightened fatigue and reduced long-term viability. Ultimately, participants did not ask for lowered standards; they asked for systems to recognize cognitive diversity and provide flexibility. 

\textbf{Over-reliance and Avoidance}.
Beyond institutional settings, participants described similar misalignments in productivity tools intended to support task management. Some expressed concern about becoming overly dependent on AI or planning systems whose affordances emphasized tracking or optimization over action. Elaborate planning sometimes became a form of avoidance rather than support, blurring the line between coping and procrastination, particularly when tools operated in isolation from social accountability, shared commitments, or situational triggers that tied plans to action. As P9 reflected, \textit{"I sometimes spend more time organizing the to-do app than doing the task."}

\subsection{Strategies: Social Dynamics as Executive Function Prosthetics}
\label{section: social strategies}
\textbf{Co-Regulation and Borrowed Structure.} Many participants described compensating for executive function challenges by "borrowing" structure from others, through body-doubling, shared routines, or friend-based monitoring systems (P2, P6, P17). These arrangements, whether formal or informal, offered external accountability that participants struggled to generate internally. P17 shared, \textit{"My friend calls me to remind me to get up and do things. Without that, I might just stay in bed all day."} Several participants deliberately built friendship networks around complementary strengths. Friends were enlisted to help manage tasks, keep schedules, or provide emotional check-ins (P4, P7). 

Participants emphasized that this support was not merely practical but deeply emotional, helping them feel less alone in their struggles. However, the fragility of these systems was also clear. P1 described a tendency to prepare for inevitable relational breakdowns by classifying friends by function and maintaining “replacements” to avoid emotional dependence, revealing the cognitive and ethical toll of over-managed interpersonal systems. As P1 explained, \textit{“I’m always worried relationships will fall apart, so I keep making new ones. When one fades, it’s painful—but I expect it.”}

\textbf{Peer Communities and Recognition.} For some, online or offline ADHD communities offered vital spaces of resonance and relief. These spaces allowed participants to express themselves without fear of judgment or having to “translate” their experiences. As P13 put it, \textit{“Only in the ADHD group do I feel that I can tell the truth.”} These communities often acted as emotional safety nets when individual or familial supports fell short.

\subsection{Strategies: Technological Tools as Externalized Cognition}

\label{section: technological tool strategies}

\textbf{Digital Tools and Scaffolded Memory.} Participants employed various tools, such as calendars, alarms, to-do lists, and habit trackers, to offload memory and enhance task initiation (P3, P5, P11, P17). Yet these tools only worked when integrated into personalized routines or embedded in relationships. P11 described sharing a calendar with a roommate for mutual follow-through: \textit{"If it’s only me reminding myself, it just becomes invisible after a few days."} Without reinforcement, tools faded into digital clutter.

\textbf{Emotional Symbolism of Tools.} For many, productivity tools served a symbolic or emotional purpose more than a functional one. The act of writing lists or maintaining journals became a ritual of containment, an attempt to impose order on internal chaos. P15 noted: \textit{"The bullet journal is the only paper tool I can stick to. It gives me peace of mind."} Similarly, P1 described list-making as pleasurable regardless of follow-through: \textit{"Writing a to-do list is a pleasure in itself."}

\textbf{Simulated Companionship and AI Co-Regulation.} Several participants used AI-based companions or gamified systems not merely to track tasks but to simulate social presence. These included asking AI chatbots to "role-play" as fictional characters, thus transforming task reporting into a shared performance (P3, P6, P9). As P9 explained, \textit{"Reporting progress to AI was helpful."} Others described motivational boosts from apps that mimicked friend-like interactions: “If a friend or even the app tells me to go to the library, I can start” (P3).

\label{section: AI expectation}

\subsection{Expectations: Imagining AI as a Relational, Adaptive Supporter}

Participants expressed aspirational visions for future AI systems that would act less like rigid managers and more like empathetic partners. These aspirations point to important unmet needs in current task-management tools. At the same time, prior research suggests that realizing such visions requires careful design choices, particularly around how emotional and contextual signals are interpreted and acted upon.

\textbf{Low-Friction and Proactive Interaction.}
Participants consistently emphasized that future AI systems should reduce interaction friction rather than introduce additional setup or cognitive burden. Many critiqued existing tools for requiring too many steps, enforcing rigid structures, or presenting cluttered interfaces that heightened anxiety instead of alleviating it (P1, P4, P14). As P4 put it, \textit{"Many apps are too complicated to set up, and I feel more anxious using them."} Reflecting this frustration, participants expressed a preference for proactive systems that surface support at the right moment, rather than requiring users to repeatedly initiate interaction, as P14 noted: \textit{"I hope the system comes to me instead of me opening it."}

\textbf{Multimodal and Emotion-Aware Interactions.}
Rather than single-modal reminders, participants hoped for multimodal, emotionally sensitive interactions, such as incorporating voice, visual cues, and vibration feedback (P1, P9, P14).
P1 imagined: \textit{"It would be best if it could ask me every day: How are you doing now? How much do you think you can accomplish today?"} P9 suggested combining sensory inputs: \textit{"Remind me with a combination of voice, vibration, and images."}
Participants viewed these features not as distractions but as supportive scaffolds for emotionally regulating task engagement. These expectations align with work in affective computing and multimodal interaction, which explores integrating cues from speech, facial expression, and physiological signals to enable more responsive systems~\cite{roemmich2023emotion, chakriswaran2019emotion}. At the same time, prior reviews note substantial variability in accuracy and generalizability, particularly across everyday contexts and diverse users~\cite{barrett2019emotional, diberardino2023anti,chakriswaran2019emotion, assunccao2022overview}. This suggests that inferring emotional states in real-world settings, especially without explicit user input, remains an open design challenge~\cite{10.1145/3706598.3713501}. Rather than treating these limitations as a barrier, prior work emphasizes emotion-aware designs that foreground reflection and self-report, prompting users to reflect on their state rather than making definitive inferences about internal affect.~\cite{10.1145/3313831.3376625, li2025introspectus, park2025reimagining, chen2023mirror}.

\textbf{Plan Negotiation and Context Sensitivity.}
Participants wanted AI systems that could dynamically adjust based on real-time emotional and cognitive states (P1, P9, P13). They envisioned systems that would suggest alternative paths rather than enforcing rigid schedules.
P09 requested: \textit{"I write the task to the AI, and it helps me break it down into ten small goals,"} highlighting the importance of decomposing tasks without triggering overwhelm. P13 added: \textit{"I hope the application can recognize my status and automatically generate an adapted plan."}
Rather than demanding compliance, participants envisioned AI as flexible collaborators that help scaffold partial successes, recalibrate expectations, and maintain psychological momentum without judgment. Emotional safety and responsive flexibility emerged as the key design desiderata for future ADHD-centered AI systems. These expectations align with work on adaptive and mixed-initiative systems, which examine how to balance system initiative with human judgment~\cite{hwang2022ai, nguyen2018believe}. However, prior studies in human–AI interaction show that when systems assume too much decision-making authority, users may feel less agency or ownership over outcomes, which can undermine engagement and trust~\cite{yan2024human, han2024teams}. Research consistently emphasizes keeping users “in the loop,” allowing them to inspect, adjust, and override system suggestions~\cite{orzikulova2024time2stop}. For adults with ADHD, whose motivation and attention often fluctuate, plan negotiation may therefore be effective when AI offers options and scaffolds decisions rather than autonomously enforcing or revising plans.

\section{Design Implications}

Building on our interview findings about the challenges adults with ADHD face in task management, emotional regulation, and tool use, we propose a set of design implications for creating more supportive AI-driven productivity systems. Importantly, these implications are not intended as a checklist, as different individuals may benefit from different combinations of them, depending on their cognitive style and situational needs. In this section, we first articulate the core difficulties participants reported, then propose design strategies that respond directly to those issues. 

\subsection{Designing for Relational Accountability Rather Than Solo Optimization}
Participants described the essential role of social scaffolding in enabling task initiation and follow-through. As documented in Sections \ref{section: social strategies}, individuals often relied on body doubling, peer check-ins, and collaborative environments to bridge executive function gaps. However, most current productivity systems conceptualize the user as a solitary agent responsible for setting and enforcing their own reminders. Static alarms and checklist systems failed to reproduce the subtle, emotional pressures created by human co-presence, often leading users to feel more isolated rather than supported.

To address this gap, future systems should simulate relational accountability, integrating conversational check-ins, progress co-tracking, and dynamic emotional acknowledgment into the user experience. Rather than operating as distant monitors, AI systems could adopt the role of attentive companions, capable of responding when users initiate tasks, drift into distraction, or complete milestones. For instance, a system might engage users with simple prompts like, “\textit{It seems like it’s been a while since you last checked in. Would you like to talk about how it’s going?}”,  thus providing gentle social pressure without invoking surveillance or shame.
This approach echoes findings from Jacobsen et al. \cite{moberg2025distributed} on the distributed nature of executive function in collaborative work settings and extends them into human-AI interaction. By framing productivity as a relational act rather than a solitary performance, designers can help users externalize willpower and sustain momentum through emotional connection, even if simulated.

\subsection{Supporting Time as Rhythm Rather Than Grid}
\label{section: design implications}
\begin{figure*}[t]
  \centering
  \includegraphics[width=\textwidth]{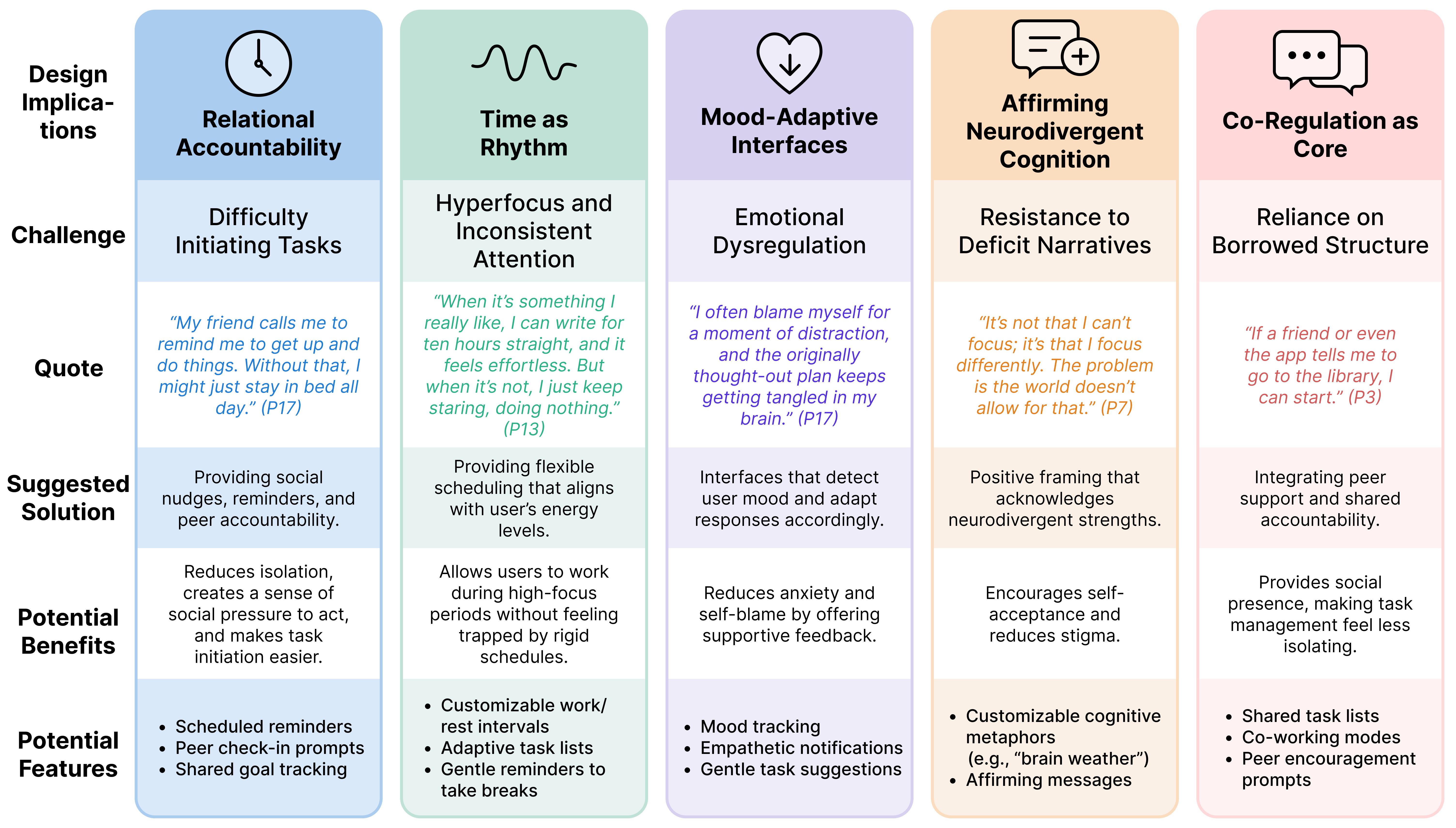}
  \caption{Design Implications Mapping}
  \label{design_implications}
\end{figure*}
A recurrent theme in Section \ref{section: internal challenges} is participants' fragmented sense of time. Many described oscillating between hyperfocus and complete inertia, with standard calendaring tools exacerbating feelings of inadequacy. Plans made during high-energy states often proved unrealistic later, leading to cycles of self-blame when rigid timelines inevitably collapsed.

Consequently, we advocate for a reconceptualization of time management interfaces. Rather than structuring tasks against fixed hours and deadlines, systems could frame time as a series of rhythms, energy flows that fluctuate naturally. This could involve supporting users in creating flexible "ideal vs. baseline" plans, where modest fallback goals are explicitly normalized alongside aspirational targets. Systems could incorporate real-time emotional and physical state check-ins, dynamically adjusting expectations when users report fatigue, stress, or emotional overwhelm.

Preserving streaks or progress markers even when users skip days could further reduce shame spirals and affirm that productivity is not linear. Prior literature on temporal structuring \cite{mazmanian2013autonomy} emphasizes the social construction of time in work systems; building flexible, rhythm-based systems would extend these insights into cognitive diversity-aware design. For adults with ADHD, temporal flexibility is not a luxury but a basic accessibility requirement.

\subsection{Developing Mood-Adaptive Interfaces to Prevent Failure Spirals}
Section 4.2 highlighted the profound influence of mood states on participants' task engagement. Negative emotions, even triggered by minor setbacks, could derail entire days, while positive emotions, if unmanaged, could lead to loss of task focus. Yet few current productivity tools acknowledge the emotional contexts that shape behavior.

We propose that future systems integrate mood-adaptive mechanisms that allow users to check in about their emotional state and adapt the system’s behavior accordingly. For instance, a user feeling overwhelmed might be encouraged to complete only a small, manageable goal, accompanied by affirming messages such as, “You seem tired, how about focusing on just one small task today?” These simple acknowledgments can prevent emotional spirals by framing reduced output as a legitimate, self-compassionate response rather than a failure. Research in affective computing and recent work on emotion-aware system design \cite{pei2024affective} reinforce the importance of tailoring system behaviors to users’ affective contexts. By embedding emotional intelligence in task management interfaces, designers can better support users’ resilience rather than inadvertently exacerbating cycles of avoidance and self-criticism.

\subsection{Affirming Neurodivergent Cognition Rather Than Pathologizing It}

While participants struggled with task initiation and planning, they also expressed pride in their distinctive cognitive patterns, particularly creativity \cite{hoogman2020creativity}, sensitivity to relevance, and associative thinking. At the same time, clinical research demonstrates substantial neuropsychological heterogeneity among adults with ADHD, indicating that ADHD does not correspond to a single cognitive style. In a large neuropsychological study, Mostert et al. \cite{mostert2018similar} identified qualitatively distinct cognitive profiles among adults with ADHD, including executive-function deficits, delay aversion, and working-memory or verbal-fluency challenges. These profiles were also present in healthy controls, suggesting that ADHD reflects performance at the extreme of normal variation. Moreover, only a minority of adults with ADHD show deficits across all cognitive domains, and high-IQ adults may compensate for executive-function difficulties through greater reasoning efficiency or learned strategies \cite{milioni2017high}.

Given this cognitive heterogeneity \cite{wolfers2020individual}, affirming neurodivergent cognition requires supporting multiple modes of engagement rather than prescribing a single ADHD-friendly interaction style. Mainstream productivity tools often pathologize divergence by enforcing uniform norms of focus, time use, and output. Future systems should instead allow users to select representations that align with their cognitive orientation. For some, this may involve ambient metaphors (e.g., “brain weather” or “focus currents”) and non-judgmental progress indicators that reflect cognitive rhythms without rigid timelines. For others, affirmation may take the form of explicit task decomposition, transparent priority logic, and adjustable constraints that appeal to more analytical or logic‑oriented users. Bennett et al. \cite{bennett2019promise} argue that inclusive design must move beyond accessibility retrofits toward affirming alternative ways of being. Our findings are aligned with this call that adults with ADHD need tools that reflect cognitive diversity \cite{Le2024Iam}.

\subsection{Embedding Co-Regulation as Core Infrastructure}
Finally, co-regulation emerged as an indispensable aspect of effective task management for our participants (Section \ref{section: social strategies} and Section \ref{section: systemic challenges}). Body doubling, accountability partners, and community validation were not secondary supports but essential mechanisms for sustaining engagement. Despite this, most productivity systems treat accountability as a matter of self-monitoring alone.

We argue that co-regulation should be built into the core architecture of future systems. Platforms could offer live "focus rooms" where users synchronize work periods with AI or human partners. Journaling dashboards could allow peer check-ins, where users report moods, progress, and obstacles to a trusted group or AI companions. Task timelines could be shared across human-AI partnerships, allowing mutual nudging and celebration of milestones without the pressures of surveillance or competition. This reimagining draws from distributed cognition theory \cite{hollan2000distributed}, which emphasizes that cognitive processes often span multiple actors and artifacts. By making co-regulation visible, accessible, and normative, systems can support users not just as isolated minds but as members of supportive cognitive ecosystems.

\section{Speed Dating} 

To explore the viability and desirability of our design implications, we conducted a speculative speed dating study using a series of design concept scenarios (See Figure \ref{tab:adhd_design_concepts}). These scenarios were developed based on our core findings and mapped directly to the design implications described in Section \ref{section: design implications}. Each scenario addresses distinct phases of the task management process, from planning and initiation to execution, adaptation, and post-task emotional recovery. This approach allowed us to probe users’ preferences, tensions, and imagined interactions with future AI-powered support systems, with each scenario grounded in challenges identified through our interviews. Importantly, while some concepts foreground individual reflection or emotional awareness (e.g., Concept 8), we conceptualize these as supporting co-regulation through simulated social presence and recognition~\cite{merrill2022ai, siemon2025beyond, walter2025designing}. By acknowledging users’ internal states and offering responsive feedback, the system creates a sense of being seen that can help stabilize motivation, reduce self-blame, and sustain engagement, even in the absence of direct human interaction~\cite{xu2025digital}.

\subsection{Design Concepts}
We developed 13 design concepts (Table \ref{tab:adhd_design_concepts}), each representing a distinct phase of the ADHD task management process: planning, task execution, daily adjustment, and posttask reflection. These scenarios were directly informed by the core challenges identified in our prior qualitative interviews and were mapped to our design implications. Each concept was crafted to explore a specific design intervention.

\begin{table*}[t]
\centering
\caption{Speed-dating design concepts and associated design implications.}
\label{tab:adhd_design_concepts}
\small
\setlength{\tabcolsep}{4pt}
\renewcommand{\arraystretch}{1.14}
\begin{tabularx}{\textwidth}{C{0.05\textwidth} L Y L}
\thickrule
\multicolumn{4}{l}{\textbf{RQ2: What design implications and speculative concept space follow for socially and affectively
aware task-management support?}}\\
\thickrule
\textbf{ID} & \textbf{Title} & \textbf{Description} & \textbf{Design Implications}\\
\midrule
1 & \textbf{Private Emotional Notes Before Planning} &
A space to jot mood notes or emotional reflections before planning; private and not used for scheduling suggestions. &
Mood-adaptive interfaces; emotional self-awareness.\\

2 & \textbf{Mood-Aware Daily Companion} &
An AI that checks in each morning, asks about feelings, and offers personalized daily suggestions. &
Relational accountability; mood-adaptive interfaces.\\

3 & \textbf{Flexible Planning and Gentle Streaks} &
Users set “ideal” and “baseline” plans; the system shifts to baseline when energy dips; streaks are preserved. &
Time as rhythm; affirming neurodivergent cognition.\\

4 & \textbf{Emotional Inventory Before New Commitments} &
Reflective prompts before accepting new tasks help evaluate energy and capacity. &
Mood-adaptive interfaces. \\

5 & \textbf{Shared Planning with a Trusted Person} &
Daily plans can be shared with a trusted person for gentle encouragement. &
Co-regulation as core; relational accountability.\\

6 & \textbf{Social Presence AI During Work} &
An AI co-worker simulates body-doubling, offering soft prompts and animated timers via visual presence. &
Co-regulation as core; relational accountability.\\

7 & \textbf{Ambient Transition Cues} &
Non-verbal ambient signals indicate focus periods or transitions. &
Mood-adaptive interfaces; affirming neurodivergent cognition.\\

8 & \textbf{Brain-Weather Visualization Dashboard} &
A dashboard visualizing cognitive states (e.g., “light fog with patches of clarity”). &
Affirming neurodivergent cognition; time as rhythm.\\

9 & \textbf{Adaptive Planning Undo Button} &
A soft “undo” lets users withdraw or rescope tasks without penalty when plans change. &
Time as rhythm; mood-adaptive interfaces.\\

10 & \textbf{Pattern-Based Gentle Nudges} &
The system detects recurring friction points and suggests adjustments. &
Time as rhythm; mood-adaptive interfaces.\\

11 & \textbf{Emotionally Aware Pause Days} &
The system suggests a pause day when signs of burnout appear and offers self-care options. &
Time as rhythm; mood-adaptive interfaces;\\

12 & \textbf{Emotional Debrief After Task Collapse} &
After abandoning a task, the system offers a non-judgmental reflection space. &
Relational accountability; mood-adaptive interfaces.\\

13 & \textbf{Weekly Narrative Reflection Instead of Analytics} &
Users narrate their week instead of viewing quantitative analytics. &
Affirming neurodivergent cognition.\\
\thickrule
\end{tabularx}
\end{table*}

For the planning stage, we propose five design concepts. First, \textbf{Concept 1: Private Emotional Notes Before Planning} offers a simple, private space where users can jot down mood notes or emotional reflections before planning their day. The system provides a text box that appears when users begin planning, but these notes are not used for any scheduling or suggestion functions. Instead, they serve purely for emotional self-awareness, helping users acknowledge how they feel before making commitments. Our findings indicated that unacknowledged emotions could derail task management (P9, P17). Second, the \textbf{Concept 2: Mood-Aware Daily Companion} is an AI that checks in each morning with a friendly greeting and asks how the user is feeling. Based on their response, it provides a personalized set of daily suggestions, balancing ambition with realism. For instance, if the user indicates low energy, it might recommend focusing on simple tasks. This design supports Relational Accountability and Mood-Adaptive Interfaces, acknowledging that emotional fluctuations impact daily performance (P2, P6, P17). Third, \textbf{Concept 3: Flexible Planning and Gentle Streaks} allows users to set two types of daily goals: an "ideal plan" (ambitious) and a "baseline plan" (minimum effort). If users experience fatigue or distractions, the system shifts to the baseline plan without resetting progress streaks. This feature directly responds to participants who reported cycles of over-planning and shame (P1, P10, P14), allowing them to maintain motivation without self-blame.

\textbf{Concept 4: Emotional Inventory Before New Commitments}, adds a reflective prompt whenever users accept a new task, such as \textit{"Do you have the energy for this right now?"} or \textit{"Would you like to schedule this for a time that suits your brain better?"} These prompts encourage conscious evaluation of emotional capacity, reducing the risk of overcommitment (P6, P17). Finally, \textbf{Concept 5: Shared Planning with a Trusted Person} allows users to share their daily plans with a trusted friend, coach, or partner. This individual can see the plan, send brief encouragements, or simply acknowledge the user's intentions without acting as a supervisor. The concept leverages Co-Regulation as Core and Relational Accountability, allowing users to feel "seen" in their efforts without pressure (P6, P13).

For the execution stage, we developed three design concepts. \textbf{Concept 6: Social Presence AI During Work} is an AI companion that simulates body-doubling, staying present with users through animated timers, gentle prompts, or simply a visual presence. Users can switch between active support (with encouraging messages) and silent companionship. This design supports Co-Regulation as Core, providing the sense of not being alone while working (P2, P6, P17). \textbf{Concept 7: Ambient Transition Cues} replace standard notifications with gentle, non-verbal signals like calming soundscapes, color shifts, or soft chimes. These cues indicate focus periods, task transitions, or breaks without creating anxiety. The concept is grounded in Mood-Adaptive Interfaces and Affirming Neurodivergence, supporting users who find traditional alarms overwhelming (P14, P4).
The \textbf{Concept 8: Brain Weather Visualization Dashboard} offers a dynamic, metaphorical display of users' cognitive states, such as "light fog with patches of clarity" or "clear skies." This dashboard helps users understand their fluctuating mental conditions, encouraging self-compassion and adaptive planning. This concept reflects Affirming Neurodivergence and Time as Rhythm (P5, P8).

Moving to the adaptation stage, we propose three more concepts. \textbf{Concept 9: Adaptive Planning Undo Button} allows users to withdraw tasks from their schedule without penalty if plans change. This feature helps users avoid planning regret and shame spirals by providing a "soft reset" without judging missed tasks (P1, P5). \textbf{Concept 10: Pattern-Based Gentle Nudges} use adaptive algorithms to detect recurring friction points (e.g., always missing morning writing sessions) and offer context-sensitive suggestions, such as \textit{"Would you like to move this task to tomorrow afternoon?"} This design supports Time as Rhythm and Self-Aware Nudging, helping users recognize and respond to their unique patterns (P10, P14). \textbf{Concept 11: Emotionally Aware Pause Days} proactively suggest a pause day if the system detects signs of burnout, offering a gentle notification like \textit{"It seems like you’ve had a tough week. Would you like to rest today?"} It also suggests self-care activities like journaling or low-pressure creative tasks (P4, P9, P17).

Finally, for after-task recovery and reflection, two concepts are designed. \textbf{Concept 12: Emotional Debrief After Task Collapse} offers a non-judgmental reflection space whenever users abandon tasks. The system asks gentle questions like \textit{"Want to reflect on what got in the way?"} helping users understand their behavior without self-blame (P4, P17).
\textbf{Concept 13: Weekly Narrative Reflection Instead of Analytics} invites users to narrate their week in their own words rather than viewing analytical graphs. It asks questions like \textit{"What surprised you? What felt unexpectedly hard?"} providing a compassionate, narrative-driven reflection experience (P13, P17).

\subsection{Participants and Procedure}

\begin{table*}[t]
\centering
\small
\setlength{\tabcolsep}{4pt}
\renewcommand{\arraystretch}{1.1}

\begin{tabular}{l c c l l c c p{3.3cm} p{3.3cm}}
\toprule
\textbf{ID} & \textbf{Gender} & \textbf{Age} & \textbf{Occupation} & \textbf{Education} & \textbf{Diag.\textsuperscript{†}} & \textbf{AI use} & \textbf{Top 3 preferred concepts (IDs)} & \textbf{Least 3 preferred concepts (IDs)}\\
\hline
S1  & M & 18 & Student         & High school   & CD  & No  & 8, 6, 3   & 9, 10, 4 \\
S2  & F & 34 & Office worker   & Bachelor’s    & CD  & No  & 6, 8, 11  & 10, 7, 5 \\
S3  & M & 23 & Student         & Bachelor’s    & SSI & Yes & 8, 11, 1  & 12, 7, 9 \\
S4  & M & 44 & Doctor          & PhD           & SSI & No  & 8, 11, 3  & 2, 5, 4 \\
S5  & F & 32 & Teacher         & PhD           & CD  & No  & 6, 8, 11  & 2, 4, 7 \\
S6  & M & 22 & Manager         & Middle school & CD  & Yes & 8, 3, 5   & 7, 4, 1 \\
S7  & F & 55 & Teacher         & PhD           & CD  & No  & 3, 6, 11  & 10, 7, 13 \\
S8  & F & 55 & Farmer          & None          & SSI & No  & 11, 12, 6 & 7, 13, 1 \\
S9  & M & 51 & Farmer          & None          & SSI & Yes & 11, 6, 8  & 7, 13, 9 \\
S10 & M & 22 & Student         & Bachelor’s    & CD  & No  & 11, 5, 7  & 9, 12, 4 \\
S11 & F & 21 & Student         & PhD           & SSI & No  & 6, 3, 9   & 4, 7, 12 \\
S12 & M & 23 & Student         & PhD           & CD  & Yes & 1, 11, 12 & 7, 4, 5 \\
S13 & F & 43 & Teacher         & PhD           & SSI & Yes & 3, 6, 8   & 5, 4, 3 \\
S14 & M & 23 & Delivery worker & Bachelor’s    & CD  & No  & 1, 6, 8   & 4, 7, 13 \\
S15 & M & 34 & Delivery worker & Master’s      & CD  & No  & 11, 8, 3  & 9, 5, 1 \\
S16 & F & 34 & Doctor          & PhD           & CD  & No  & 3, 8, 12  & 7, 4, 13 \\
S17 & M & 33 & Courier         & Bachelor’s    & CD  & No  & 11, 6, 3  & 12, 4, 9 \\
S18 & F & 21 & Student         & Bachelor’s    & SSI & No  & 3, 5, 8   & 7, 10, 6 \\
S19 & F & 21 & Student         & Bachelor’s    & CD  & Yes & 6, 11, 8  & 4, 10, 12 \\
S20 & M & 21 & Student         & Bachelor’s    & SSI & Yes & 3, 6, 8   & 7, 9, 4 \\
\hline\hline
\end{tabular}

\caption{Demographic information of speed-dating participants.}
\label{tab:speed_dating_demographics}

\par\smallskip
{\small \emph{Note.} \textsuperscript{†}\,Diag.\ abbreviations: CD = clinically diagnosed; SSI = strongly self-identified. Totals clinically diagnosed: 12; Total strongly self-identified: 8; AI use (Yes): 7; (No): 13. “Top 3”/“Least 3” = each participant’s three most/least preferred speed-dating concepts; numbers are concept IDs
(see Table~\ref{tab:adhd_design_concepts}).}
\end{table*}

\begin{figure*}[t]
  \centering
  \includegraphics[width=\textwidth]{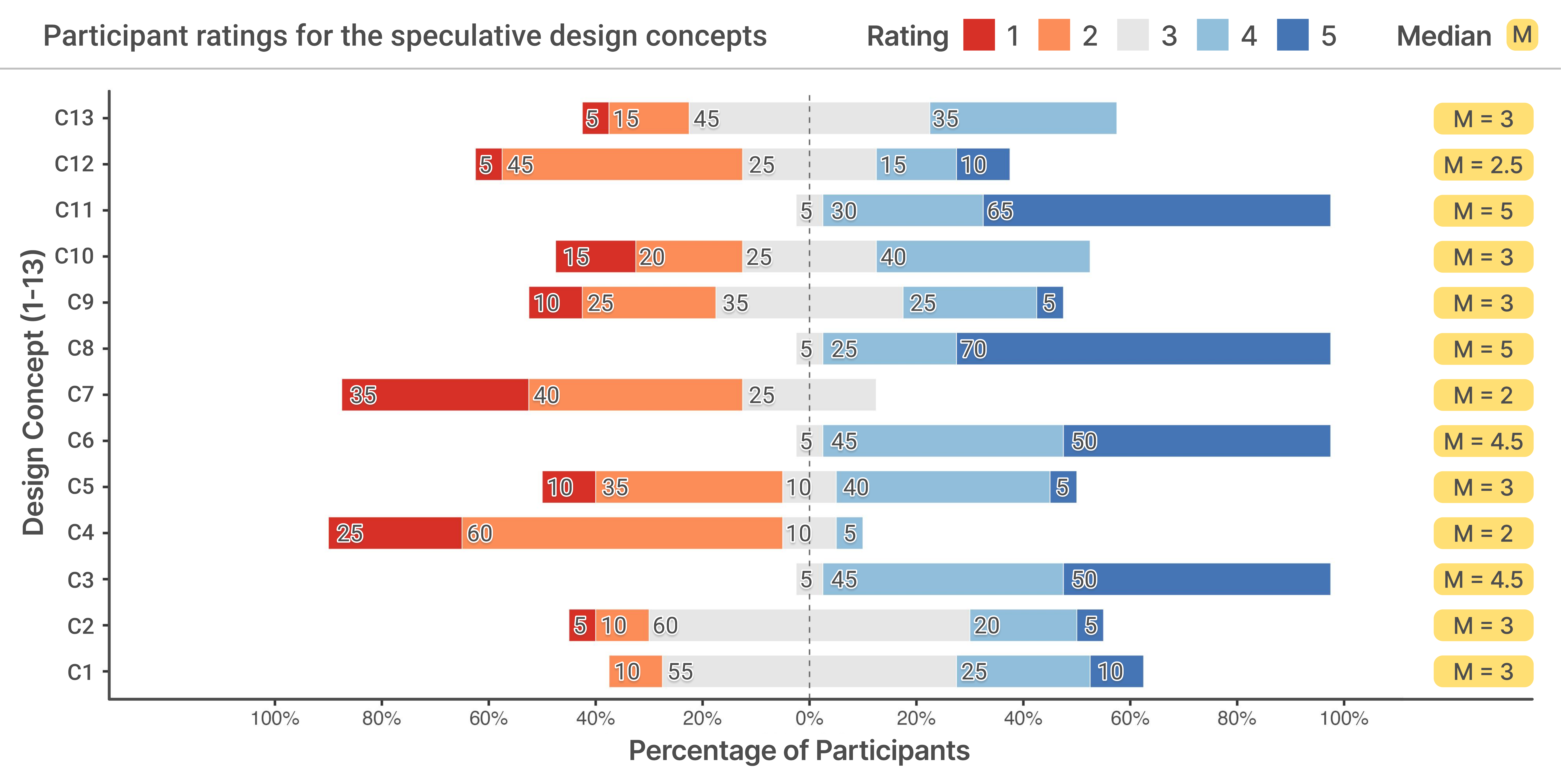}
  \caption{Overall participant ratings for the 13 speculative design concepts (N=20; 1–5 Likert scale where 1 = least preferred / would not want to use and 5 = most preferred
/ would very much want to use). Concept IDs match Table~\ref{tab:adhd_design_concepts}.}
  \label{concept}
\end{figure*}

We recruited 20 adults diagnosed with ADHD (IDs S1-S20) who did not participate in our previous interviews (Table \ref{tab:speed_dating_demographics}). Participants were recruited through online advertisements, peer referrals, and community organizations. Each session was conducted as a semi-structured interview lasting approximately 35 minutes with a 5-minute break. Participants were presented with each of the 13 scenarios in a randomized order. For each scenario, participants first reviewed the descriptions and received a brief verbal explanation, then rated the scenario on a Likert scale (1–5, where 1 = least preferred / would not want to use and 5 = most preferred / would very much want to use), and provided qualitative feedback, discussing their emotional reactions, perceived value, and concerns (See Figure \ref{concept}).
Following Davidoff et al.\cite{davidoff2007rapidly}, we used thematic coding to analyze participant responses across scenarios. Codes captured emotional tone (e.g., comfort, anxiety, shame relief), perceived fit with ADHD reality, trust, control, and relational interpretation of the AI, as well as emergent user needs or rejections. Two researchers independently coded transcripts, then met to reconcile interpretations.

\subsection{Speed Dating Findings}

Participants expressed diverse reactions to the speculative design concepts, reflecting differences in emotional needs, task contexts, and coping strategies. Rather than converging on a single preferred solution, feedback showed that different designs were useful under different circumstances.

\textbf{Concept 8: Brain Weather Visualization Dashboard} was frequently discussed in terms of emotional understanding and reduced self-blame. Participants appreciated metaphorical representations of cognitive states (e.g., “light fog,” “clear skies”) as a way to externalize fluctuations without stigma. Several contrasted this with conventional productivity indicators such as overdue task lists. S5 stated, “\textit{I hope my emotions are seen, I hope I am seen and understood},” while S9 described “mild haze” as more comforting than red overdue tasks, characterizing it as poetic and calming. Participants also noted that adapting task recommendations to perceived state helped sustain a sense of accomplishment during low-energy periods.

\textbf{Concept 6: Social Presence AI at Work} prompted responses centered on companionship and reduced isolation. Participants likened the AI to a “body double” offering quiet presence rather than oversight. S2 described it as a partner who “understands and supports and doesn’t judge,” while S6 emphasized its usefulness when human companionship was unavailable. S7 compared the AI to the rose in \textit{The Little Prince}, valuing emotional reassurance over direct productivity gains.

\textbf{Concept 11: Emotionally Aware Pause Days} surfaced participants’ difficulty recognizing emotional overload early. Many valued a system that framed rest as care rather than failure. S1 noted that sadness often became apparent only when overwhelming, and S17 suggested that earlier intervention might have prevented burnout. Others raised concerns about misinterpretation, emphasizing transparency and user control.

\textbf{Concept 3: Flexible Planning and Gentle Streaks} resonated with participants who struggled with self-criticism around incomplete plans. The dual “ideal” and “baseline” goals were described as acknowledging effort without abandoning ambition. S14 said this stopped them from “hating themselves for not completing everything,” while S19 noted that flexibility made attention feel developable over time. Some cautioned that excessive flexibility could become demotivating, highlighting tensions between self-compassion and accountability.

Reactions to other concepts were more situational. \textbf{Concept 5: Shared Planning with Trusted Persons} elicited mixed responses: some valued gentle accountability (S9), while others found it intrusive or emotionally demanding. S4 questioned whether supportive relationships between family or friends should require AI mediation. \textbf{Concept 10: Pattern-Based Gentle Nudges} similarly divided participants. Some appreciated nudges that aligned tasks with personal rhythms, while others found them intrusive or ineffective. S2 felt that the system acknowledged emotions without materially helping, whereas S14 suggested that
emotional care could still be valuable even when efficiency declined.

Some concepts were perceived as less aligned with participants’ attentional states. \textbf{Concept 7: Ambient Transition Cues} were often too subtle to notice during focused work; several participants reported they are likely to miss them entirely. Others suggested customization or reframing them for mood awareness rather than task transitions. \textbf{Concept 4: Emotional Inventory Before New Commitments} prompted mixed responses. Some valued it as a private outlet for emotional expression, while others felt it introduced friction during moments of high motivation or impulsive task acceptance.

Additional concepts illustrated similar tradeoffs. \textbf{Concept 1: Private Emotional Notes} was seen as useful for emotional expression but unnecessary within task-focused workflows. \textbf{Concept 9: Elastic Undo Button} reduced anxiety for some (S11) but raised concerns about enabling procrastination. \textbf{Concept 12: Emotional Review of Task Failure} supported reflection for some (S16, S5) while others worried about revisiting failure too directly. \textbf{Concept 13: Narrative Summary} was generally seen
as useful for sensemaking but not distinctive, suggesting that narrative reflection alone may be insufficient without additional
scaffolding.

Taken together, the findings indicate that adults with ADHD do not benefit from a uniform model of task-management support. Participants valued different combinations of emotional acknowledgment, flexibility, reflection, and social presence depending on context and preference. Designs emphasizing emotional understanding (Concept 8), adaptive rest (Concept 11), and supportive presence (Concept 6) were often helpful during periods of low energy or strain. In contrast, concepts perceived as surveillance-like (Concept 5) or cognitively burdensome (Concept 4) were met with resistance. Across scenarios, participants emphasized user control: adaptive support was most welcome when users could configure nudges (Concept 10), adjust cue salience (Concept 7), or opt in and out as needs changed.

\section{Discussion}

\subsection{Socially and Emotionally Scaffolded Task Management}

Our analysis shows that adults with ADHD rarely manage tasks in isolation. Instead, task management emerges as a socially and emotionally scaffolded process, in which participants rely on body doubling, peer accountability, adaptive tools, and affective support to initiate and sustain action. These practices transform task management from an individual act of self-discipline into a cooperative and relational accomplishment, where progress depends on the availability and stability of external supports.

While this pattern resonates with theories of distributed cognition \cite{hollan2000distributed} and cognitive offloading \cite{kirsh2010thinking}, our findings suggest that scaffolds were not simply informational or instrumental but emotionally contingent: participants’ ability to engage with tasks depended on feeling understood, encouraged, or non-judged. Emotional fluctuations, such as anxiety, shame, or sudden loss of motivation, could rapidly destabilize otherwise effective support structures, causing participants to withdraw from tools, plans, or collaborators altogether. This highlights a form of distributed regulation that is fragile and situational. By foregrounding emotional co-regulation, our findings extend prior work on socially scaffolded productivity \cite{pea2018social} and affective computing \cite{picard2000affective}. Rather than treating emotion as a secondary layer that modulates task execution, our data suggest that emotional regulation is integral to whether task scaffolds function for adults with ADHD.

\subsection{Design Conflicts and Implications in creating ADHD-supportive AI systems}

Our analysis of the speed dating results reveals several critical design conflicts in ADHD-supportive AI systems, highlighting the nuanced and often contradictory needs of ADHD users. These conflicts primarily revolve around four key areas: autonomy versus guidance, emotional support versus efficiency, privacy versus social connection, and adaptive nudges versus intrusive reminders. These tensions not only reflect the diverse experiences of ADHD users but also point to the inherent complexity of designing AI systems that can genuinely support neurodivergent individuals.

One of the most prominent tensions emerged around the balance between adaptive AI guidance and user autonomy. Participants expressed a clear preference for tools that provide suggestions rather than directives. For instance, while many valued adaptive recommendations that aligned with their mood and energy levels, as seen in the highly rated Brain Weather Dashboard (Concept 8), others feared excessive intervention or a sense of being micromanaged. As one participant noted, "I want it to suggest, not decide." This feedback underscores the importance of designing AI systems that offer context-sensitive nudges without overriding user control \cite{oz2023role, hwang2022ai}. 

Another significant conflict was between emotional support and efficiency. Participants consistently valued designs that acknowledged their emotional state, reduced self-blame, and validated users' feelings without imposing rigid expectations. Such designs align with work on emotion regulation intervention for on-the-spot intervention, emphasizing emotion regulation strategies during naturally occurring emotional situations \cite{Slovak2023Emotion}. An opposite example would be Concept 4: Private Emotional Notes Before Planning, which was seen as burdensome. Many users viewed emotional journaling as an extra step that disrupted their workflow rather than enhancing it. This divergence reveals a fundamental insight: while emotional validation is critical, it must be seamlessly integrated into user interactions without creating additional cognitive or emotional burden. One promising approach to address this challenge is the implementation of Just-In-Time Adaptive Interventions (JITAIs)\cite{nahum2018just, hardeman2019systematic, juarascio2018just, chen2024situfont}. JITAIs are designed to provide the right type and amount of support at the right time by adapting to an individual's changing internal and contextual state \cite{nahum2023adapting}. By leveraging real-time data, such as mood or productivity levels, JITAIs can deliver interventions precisely when users are most receptive, thereby minimizing disruption and enhancing effectiveness \cite{nahum2015building, gonul2019expandable}. 

The tension between privacy and social connection also emerged as a core design challenge. Some participants found comfort in the simulated social presence provided by Concept 6: Social Presence AI at Work, which mimicked the experience of a "body double" and offered a sense of shared presence without the anxiety of direct supervision. In contrast, Concept 5: Shared Planning with Trusted Persons was divisive. For some, it provided gentle accountability and strengthened social connections, while others viewed it as an invasion of privacy or a source of relational pressure. This reflects a broader challenge in designing for ADHD users: the fine line between providing supportive social connections and avoiding unwanted social obligations \cite{carpenter2009social, michielsen2015relationships, zhang2025towards}.

Finally, adaptive nudges also generated mixed responses. Concept 10: Pattern-Based Gentle Nudges aimed to provide context-sensitive reminders aligned with users' natural rhythms. While some participants appreciated the gentle encouragement, others found these nudges intrusive, with S14 noting, \textit{"Even if I know it’s trying to help, it feels like it’s watching me."} This conflict emphasizes the need for customizable AI systems where users can control the frequency, tone, and content of reminders. Adaptive behavioral support must be user-directed rather than feeling like constant surveillance~\cite{ferreyra2020persuasion}.

These design conflicts illuminate the fundamental challenge in creating ADHD-supportive AI systems: striking a balance between being adaptive without becoming overbearing, supportive without being rigid, and socially aware without compromising privacy.

\subsection{Ethical Considerations: Simulated Companionship and Autonomy}
AI systems offer a unique advantage for ADHD users by providing consistent, non-judgmental support. Participants frequently emphasized that, unlike human interactions, AI systems did not impose expectations, criticize, or stigmatize their behaviors. This aligns with existing research on human-technology interaction \cite{turkle2011alone}, which suggests that non-judgmental AI can reduce user anxiety and create a sense of emotional safety \cite{hu2025ai, meng2023mediated}. For adults with ADHD, who often experience social misrecognition and invalidation \cite{kooij2012adult}, the ability to engage with a supportive system without fear of judgment is strongly desired.

However, this benefit is double-edged. Although non-judgmental support can foster self-acceptance, it may also lead to over-reliance on AI systems, especially if users come to prefer synthetic support over authentic social interactions, especially under anxiety for social obligations \cite{hu2023social}. This tension echoes ethical concerns in HCI around fear of simulated companionship in replacing genuine human relationships \cite{ciriello2024ethical}. While AI offers emotional consistency, it risks creating an emotional echo chamber \cite{mlonyeni2025personal}, where users receive validation without challenge. Our findings advocate for a balanced approach, where AI provides supportive, non-judgmental interactions while encouraging users to maintain and develop authentic social connections.

\subsection{Diagnosis as Sense-Making and Legitimacy}

In reflecting on participants’ accounts across interviews and the speed-dating study, we found that diagnostic status primarily shaped how ADHD experiences were interpreted rather than the challenges themselves. Clinically diagnosed (CD) participants often described diagnosis as a form of retrospective re-interpretation, offering emotional closure and coherence by reframing earlier struggles that had previously been interpreted as personal shortcomings through the lens of neurodivergence. For instance, P4 described their diagnosis as "forgiveness". In contrast, strongly self-identified (SSI) participants more frequently described ADHD as a present-tense, lived friction, marked by ongoing negotiation with attention, motivation, and emotional regulation. Despite these narrative differences, both groups reported largely overlapping task-management challenges and AI design preferences, with no systematic differences observed in the speed-dating study (see Table \ref{tab:speed_dating_demographics}). However, because our sample was not balanced across diagnostic categories and we did not independently validate diagnostic status, we cannot reliably compare experiences across clinically diagnosed and strongly self-identified participants or draw subgroup-level conclusions. 

\section{Limitations}

This study has several limitations. First, our sample likely skews toward introspective, self-motivated, and tool-using adults with ADHD. Recruitment through social media and community referrals may have over-represented people who are already reflective about their ADHD experiences and comfortable experimenting with digital tools, while under-representing those with lower technology access, less diagnostic literacy, or limited time for research participation. As such, the findings may not fully reflect the heterogeneity of the broader ADHD population.

Second, our inquiry was cross-sectional and relied on relatively short sessions. While these formats surfaced rich narratives and reactions, they cannot capture how practices stabilize (or unravel) over weeks and months, nor how preferences shift as life circumstances, supports, or symptoms change. In addition, our methods are inherently reliant on participants' self-reported experiences and perceptions. This introduces the potential for recall bias, social desirability bias, and subjective interpretations, which may not always accurately reflect participants' real-world behaviors or challenges.

Third, approximately one third of participants across our studies self-identified as having ADHD without a formal medical diagnosis. The inclusion of self-diagnosed participants was a deliberate methodological choice, considering structural barriers to adult ADHD diagnosis such as prolonged assessment delays, financial costs, gendered diagnostic biases, and the historical under-recognition of adult ADHD~\cite{elefante2025raising, rivas2023adult}. However, because we did not independently validate diagnostic status or assess symptom severity using standardized screening instruments, we cannot reliably compare experiences across clinically verified subgroups (inattentive, hyperactive-impulsive, combined), medication status, or common co-occurring conditions. Consequently, subtype-specific challenges and supports may have been obscured, and our findings should be interpreted as characterizing ADHD-identifying adults rather than diagnostically stratified populations.

Fourth, the AI concepts were evaluated through speculative, researcher-led speed-dating rather than through in-situ, working systems. This approach was appropriate for mapping a wide design space, but it lacks ecological validity: real-world use would surface additional factors such as UI friction, reliability, latency, privacy boundaries, breakdown recovery, and trust calibration that can materially shape adoption and outcomes. In addition, responding to pre-designed scenarios may have constrained participants’ creativity relative to more open-ended co-design.

\subsection{Future Work} Future work should undertake longitudinal, in-situ deployments of minimally viable prototypes that instantiate our socially and affectively scaffolded designs to observe how practices stabilize or unravel outside the lab. To relate moment-to-moment affect and attention to uptake and outcomes, studies can combine diary measures with unobtrusive instrumentation; importantly, outcomes should extend beyond completion rates to include task initiation, co-regulation, shame reduction, and attentional stability. In addition, sampling should be broadened beyond digitally engaged participants and stratified by ADHD presentation (inattentive, hyperactive-impulsive, combined), medication status, common co-occurrences, and cultural context, thereby enabling subtype-aware and cross-cultural analyses. Because Adult ADHD presentations reflect predominant symptom clusters rather than stable traits, and often fluctuate across contexts, treatment phases, and the life course~\cite{willcutt2012prevalence, gibbins2010adhd}, longitudinal designs are necessary to capture within-person shifts in attention, motivation, and regulation that static categorizations cannot explain. Incorporating validated screening instruments (e.g., the Adult ADHD Self-Report Scale) alongside self-identification would enable subtype-aware analyses and allow future work to examine how formal diagnosis, treatment access, and support infrastructures shape task management practices and engagement with AI-based scaffolds over time. Field evaluations should foreground appropriation and breakdowns, boundary-setting around privacy and trust, and risks of over-reliance on simulated companionship. Finally, comparative studies with other cognitively diverse populations can test where these design principles generalize to CSCW systems and where divergence is warranted.

\section{Conclusion}
In this study, we explored the challenges of task management among adults with ADHD, uncovering how their task management practices are not solely cognitive but are deeply rooted in emotional, social, and relational contexts. Our research, which began with 22 in-depth interviews followed by a speed dating study involving 20 additional participants, revealed that task management for ADHD adults is a socially and emotionally scaffolded practice rather than an isolated individual endeavor.
Thus, AI-augmented social scaffolds can be useful in providing adaptive, relational support that aligns with users' emotional states and cognitive rhythms. Rather than enforcing rigid schedules or linear productivity metrics, such systems can dynamically adjust to user needs, offering gentle nudges, emotional affirmations, and context-sensitive planning tools. By treating task management as a cooperative and emotionally-aware process, our approach highlights the potential for AI to serve not just as a productivity monitor but as an empathetic partner for ADHD adults.

\bibliographystyle{ACM-Reference-Format}
\bibliography{sample-base}

\appendix
\section{Appendix: Semi-Structured Interview Protocol}

\textbf{Participant Background and Daily Context)}
\begin{itemize}
\item Basic self-introduction (age, occupation, education).
\item Can you describe your daily life? (Work, study, etc.)
\item How do you perceive your ADHD? What are its specific manifestations in your daily life?
\item Are there specific times or situations where you feel the impact of ADHD is more pronounced?
\end{itemize}

\textbf{Overall Productivity Strategies}
\begin{itemize}
\item How do you define an "efficient day"? What does being productive mean to you?
\item What obstacles do you typically encounter in improving your productivity?
\item Do you have any personal strategies to help you stay focused or complete tasks?
\item Have these strategies changed over time?
\item Are there specific times or rhythms when you feel most productive?
\end{itemize}

\textbf{Task Management Exploration}
\begin{itemize}
\item How do you currently manage your tasks? (Apps, paper lists, memory, calendar, etc.)
\item What aspects of your current system work well? What aspects are frustrating or ineffective?
\item What types of tasks are hardest for you to start? Why?
\item How do you usually get yourself to begin?
\item Can you share an example where you struggled to start something?
\item How do you feel when you need to switch from one task to another?
\item Do you ever find it hard to stop a task once you start?
\item What methods or tools help you transition between tasks?
\item What are your most significant sources of distraction?
\item Do you manage different types of tasks differently? (Creative vs. routine, deadline vs. no deadline, interesting vs. uninteresting)
\item Do you have others who assist with your task management? (Friends, coaches, colleagues)
\item Have you tried “body doubling” (working alongside someone else, either physically or virtually)?
\end{itemize}

\textbf{Imagining and Attitudes Toward AI Tools}
\begin{itemize}
\item What would your ideal task assistant look like?
\item What kinds of support would you find helpful? What features might feel intrusive or useless?
\item If the system’s appearance or communication style could be personalized, what would you prefer?
\end{itemize}

\textbf{Conclusion and Appreciation}
\begin{itemize}
\item Is there anything we haven’t asked that you think is important?
\item If someone is designing a productivity tool for ADHD individuals, what advice would you give?
\end{itemize}



\end{document}